\documentclass[12pt]{iopart}

\usepackage{iopams,graphicx}
\begin{document}

\title[Mechanisms for photon sorting based on slit-groove arrays]{Mechanisms for photon sorting based on slit-groove arrays}

\author{F. Villate-Gu\'io$^1$, L. Mart\'in-Moreno$^1$ and F. de Le\'on-P\'erez$^{1,2}$}

\address{$^1$Instituto de Ciencia de Materiales de Arag\'on and Departamento de F\'isica de la Materia Condensada, CSIC-Universidad de Zaragoza, E-50009 Zaragoza, Spain}
\address{$^2$Centro Universitario de la Defensa de Zaragoza, Ctra. de Huesca s/n	, E-50090 Zaragoza, Spain}
\ead{fdlp@unizar.es}
\begin{abstract}
Mechanisms for one-dimensional photon sorting  are theoretically studied  in the framework of a  couple mode method. The considered system is a nanopatterned structure composed of two different pixels drilled on the surface of a thin gold layer. Each pixel consists of a slit-groove array designed to squeeze a large fraction of the incident light into the central slit. The Double-Pixel is optimized to resolve two different frequencies in the near infrared. This system shows a high transmission efficiency and a small crosstalk. Its response is found to strongly depend on the effective area shared by overlapping pixels. Three different regimes for the process of photon sorting are identified and the main physical trends underneath in such regimes are unveiled. Optimal efficiencies for the photon sorting are obtained for a moderate number of grooves that overlap with grooves of the neighbor pixel. Results could be applied to optical and infrared detectors.

\end{abstract}

\maketitle

\section{Introduction}
Coupling between electromagnetic fields and surface modes in patterned metallic nanolayers offers the possibility for new mechanisms to guide, trap and localize light \cite{FJRMP10}. The optical response of nanostructured metallic layers is characterized by narrow spectral bands with resonant wavelengths mainly determined by the periodicity of the structure. Therefore, these systems can be used as filters by just tuning the periodicity \cite{GenetN07}. A full analysis of the dependence of such resonances on other geometrical parameters is also available in the literature for systems like hole arrays \cite{EbbesenN98} and apertures surrounded by corrugations \cite{ThioOL01,LezecS02,HibbinsAPL02,FJPRL03,AkarcaAPL04,ThomasSSC04,JanssenPRL07}.

In technological applications, like digital cameras or displays, color discrimination is performed through arrays of pixels, where each pixel acts as a separate entity sensitive to a single color \cite{Bayer}. Multispectral sensitivity have been also demonstrated in systems like waveguide resonators \cite{KangOE08,DiestNL09,XuNCom10} and light harvesting structures, as for example in triangular arrays of apertures surrounded by corrugations \cite{LauxNP08} or in a mosaic of free-standing arrays of slits used as band pass filter\cite{haidarAPL10}. Such arrangements of nanostructured metallic pixels with multiple spectral resonances behave as wavelength-selective devices with promising advantages in spacial resolution. Furthermore, this allows the extension of reliable filtering into regions of the spectrum other than the optical.  

In addition to their capability for selecting frequencies,  these devices are also able to guide photons with different wavelengths through different channels, i.e. they can be considered as photon sorters; see, for instance, the overlapping light-collection structures reported in Ref. \cite{LauxNP08}, where each pixel is devoted to harvest light of a single color and squeeze it through the central aperture. This device acts as a spectrometer that detects different wavelengths in the same area. Simultaneously, it can be used to generate an image of the object, fulfilling in this way the requirements of the spectral imaging methodology \cite{GariniCyt06}. Laux et al. have proposed in Ref. \cite{LauxNP08} both 2D (bull's eye structures) and 1D (slit-groove arrays) versions of such photon sorters.

In this paper, we first study the main physical mechanisms appearing in the process of photon sorting and, second, how to optimize the 1D version of such photon sorters in order to render highly efficient light-harvesting processes. 

\begin{figure}[h]
\centering
\includegraphics[width=8.5cm,height=3cm]{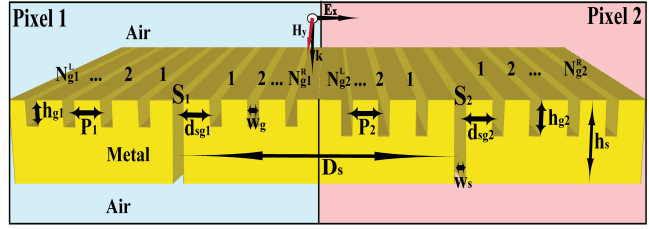}
\caption{(Color). Schematic representation of the Double-Pixel. Both pixels are sculptured on a uniform gold layer with thickness $h_s$. Pixel 1 has a central slit $S_1$ of width $w_{s_1}$ surrounded by grooves of periodicity $P_1$, depth $h_{g_1}$ and width $w_{g_1}$. The distance between the slit and the nearest groove is $d_{sg_1}$. Similar parameters are defined for pixel 2. The distance between slits is $D_s$.}
\label{fig:scheme}
\end{figure}

The building block of the structure considered in our work is a thin gold layer (optically opaque) perforated with a subwavelength slit, which is surrounded by an array of periodic grooves sculpted on the illuminated surface, see Fig \ref{fig:scheme}. Geometrical parameters of this slit-groove array (SGA) are adjusted in order to make the system resonant to a given wavelength. Optimal geometries can be obtained following simple rules recently reported \cite{FJPRL03,JanssenPRL07,VillateOE12}. This system can be easily integrated with a standard photodetector \cite{CollinAPL03,IshiJJAPL05364,YuAPL06,DunbarAPL09,BeriniLPR14}, sensitive to the narrow band of the resonant wavelength. 

Two isolated SGAs, with optimal response at  targeted wavelengths $\lambda_1$ and $\lambda_2$, are designed. The SGAs are arranged  in a Double-Pixel, as shown in Fig \ref{fig:scheme}, that resolves both wavelengths with a small cross-talk between the pixels. As we are interested in efficient mechanisms to collect the light impinging on a given pixel and redirect it to the other, i.e. to design and implement a photon sorter, we study the change in the optical response of the Double-Pixel as a function of the overlap between the SGAs. 

We focus our attention on the near-infrared part of the spectrum. Integrating light harvesting structures on IR detectors has been recently proposed as an efficient way to increase the absorption of light in a given volume \cite{RossNP12}. In this way it is possible to reduce the noise and raise the output signal. Results obtained here can be easily extended to other parts of the IR spectrum as well as to optical frequencies. 

The paper is organized as follows. Next section describes the theoretical framework. Our results are discussed in Sec. \ref{sec:mechanisms}. Section \ref{sec:sensing} shows how to use non-overlapping pixels as a sensing device. Section \ref{sec:sorting} studies in detail the influence of the overlap between pixels, paying attention to physical mechanisms for photon sorting. The influence of the number of grooves in the optical response of overlapping pixels is analyzed in Sec. \ref{sec:number}. At the end of the paper our main conclusions are given.

\section{Theoretical framework}
An schematic representation of the Double-Pixel is given in Fig. \ref{fig:scheme}. Both pixels are sculptured on a uniform gold layer with thickness $h_s$. The dielectric constant of gold is taken from Ref. \cite{Palik}. Pixel 1 has a central slit $S_1$ of width $w_{s_1}$ surrounded by grooves of periodicity $P_1$, depth $h_{g_1}$ and width $w_{g_1}$. The distance between the slit and the nearest groove is $d_{sg_1}$. Similar parameters are defined for pixel 2. The distance between slits is $D_s$. The system is illuminated by a plane wave incident perpendicular to the metal surface, with its electric field parallel to the x axis. 

The number of grooves at either side of the slit can be in principle different. So, we have $N^L_{g_1}$ grooves sculpted on the left side and $N^R_{g_1}$ on the right side of pixel 1; while pixel 2 has $N^L_{g_2}$ and $N^R_{g_2}$ grooves on the left and right sides of the slit, respectively. We use $N^L_{g_1}=N^R_{g_2}=6$ along the paper. Only the number of grooves located between the two slits ($N^R_{g_1}$ and $N^L_{g_2}$) are changed.

Calculations are done in the framework of the coupled-mode method (CMM) \cite{FJRMP10}. This semi-analytical approach nicely reproduces experimental results on SGAs \cite{DunbarAPL09,FLTNP07,FLTNJP08}. 

The CMM is based on a convenient representation of the EM fields. Above and below the metal film the fields are expanded into an infinite set of plane waves with both p- and s-polarizations. Inside slit and grooves the most natural basis is a set of planar waveguide modes \cite{stratton}. Convergence is fast achieved with a small number of such waveguide modes. The parallel components of the fields are matched at the metal/air interface using surface impedance boundary conditions \cite{jackson}. These boundary conditions are also applied at the lateral walls of slit and grooves \cite{LochbihlerJMO93}. After matching the fields at the interface we arrive to a linear system of tight binding-like equations that can be easily solved \cite{FJPRL03,FdLPNJP08}. 

Using the CMM, we compute the normalized-to-area transmittance ($\eta$), which is defined as the intensity of the light radiated to the far-field normalized to the intensity of the light incident on the area of the slits. It accounts for the efficiency of the light harvesting process: $\eta$ is of the order of 1 for a single slit, whereas it could become one or two orders of magnitude larger when the groove array squeezes additional light to the central slit \cite{FJPRL03,JanssenPRL07,VillateOE12}. 

\section{Mechanisms for the photon sorting}
\label{sec:mechanisms}
\subsection{Spectral response}
\label{sec:sensing}
The largest transmittance for an isolated SGA is obtained when the Fabry-Perot mode of the slit is located at the same spectral position of the groove cavity mode \cite{FJPRL03}. For a given wavelength, the spectral position of Fabry-Perot mode can be tuned by both metal thickness $h_s$ and slit width $w_s$, while the groove cavity mode of a groove array is  a function of the groove depth and width \cite{FJPRL03}. The optimal periodicity should guarantee that all light re-emitted from the grooves reach the other grooves and the central slit in phase. Varying the distance from the slit to the its nearest groove allows a further control of the interaction between the slit and the groove array. Groove pitch and depth are the most relevant design parameters. Ideal values of $w_g$ and $d_{sg}$ allow a finer tuning of the transmittance. Detailed design rules have been reported in a previous work \cite{VillateOE12}.

As a proof of principle, pixels 1 and 2 are designed to operate at $\lambda_1=1.35$ $\mu$m and $\lambda_2=1.50$ $\mu$m, respectively. A typical experimental value of $w_s=100$ nm is chosen for the slit width. Both slits have the same width for the sake of simplicity. We use a constant metal thickness $h_s=390$ nm, which is the arithmetic mean of the optimal thickness values needed to excite the Fabry-Perot modes in $S_1$ and $S_2$. This uniform layer does not favor any particular Fabry-Perot mode. Both groove arrays have 6 grooves at each side of the slit. Ref. \cite{VillateOE12} shows that well-defined and high-intensity transmission peaks with a full-width half-maximum (FWHM) of the order of 100 nm are obtained for a SGA with 12 grooves.   

\begin{figure}[h]
\centering
\includegraphics[width=8cm]{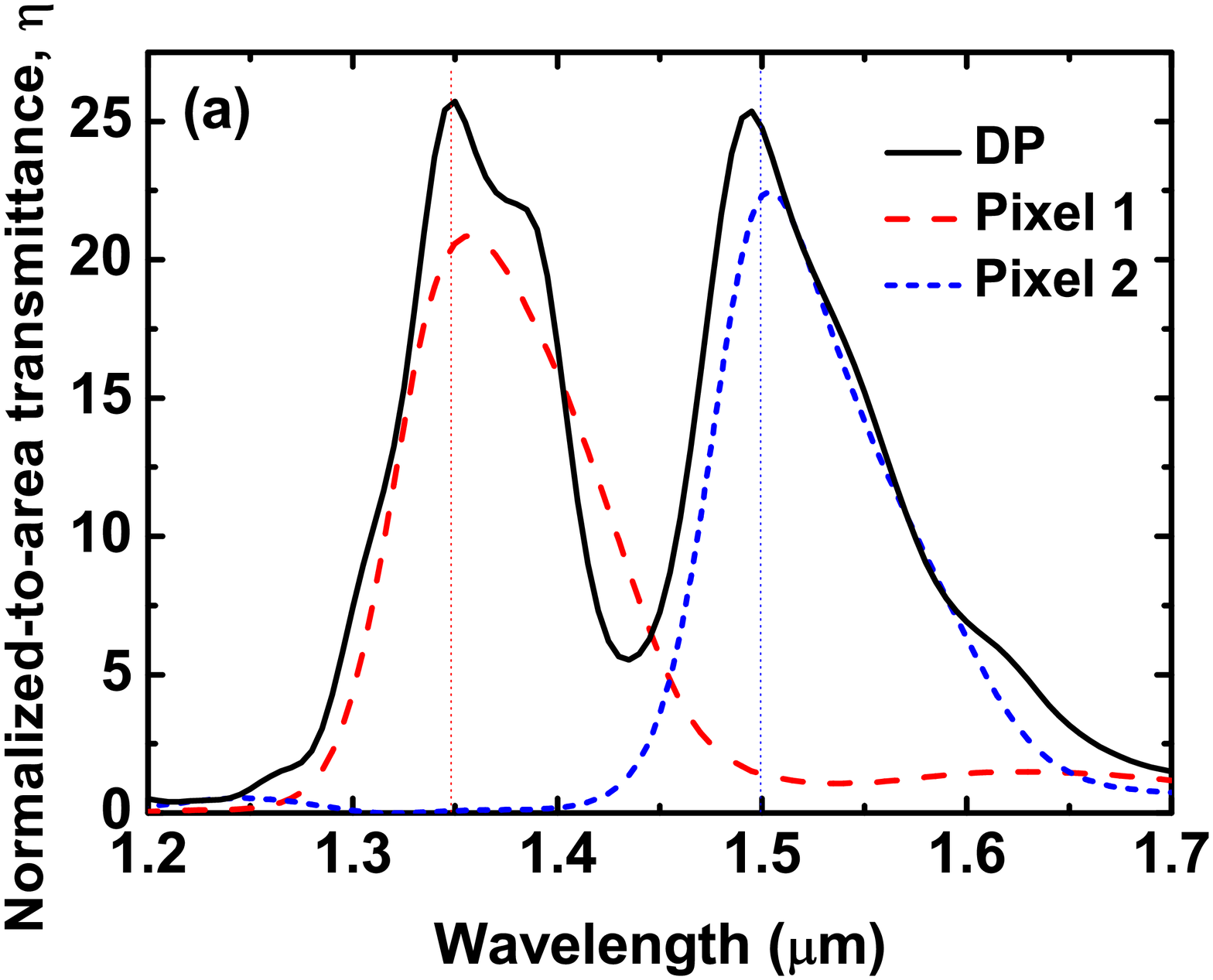}
\includegraphics[width=8cm]{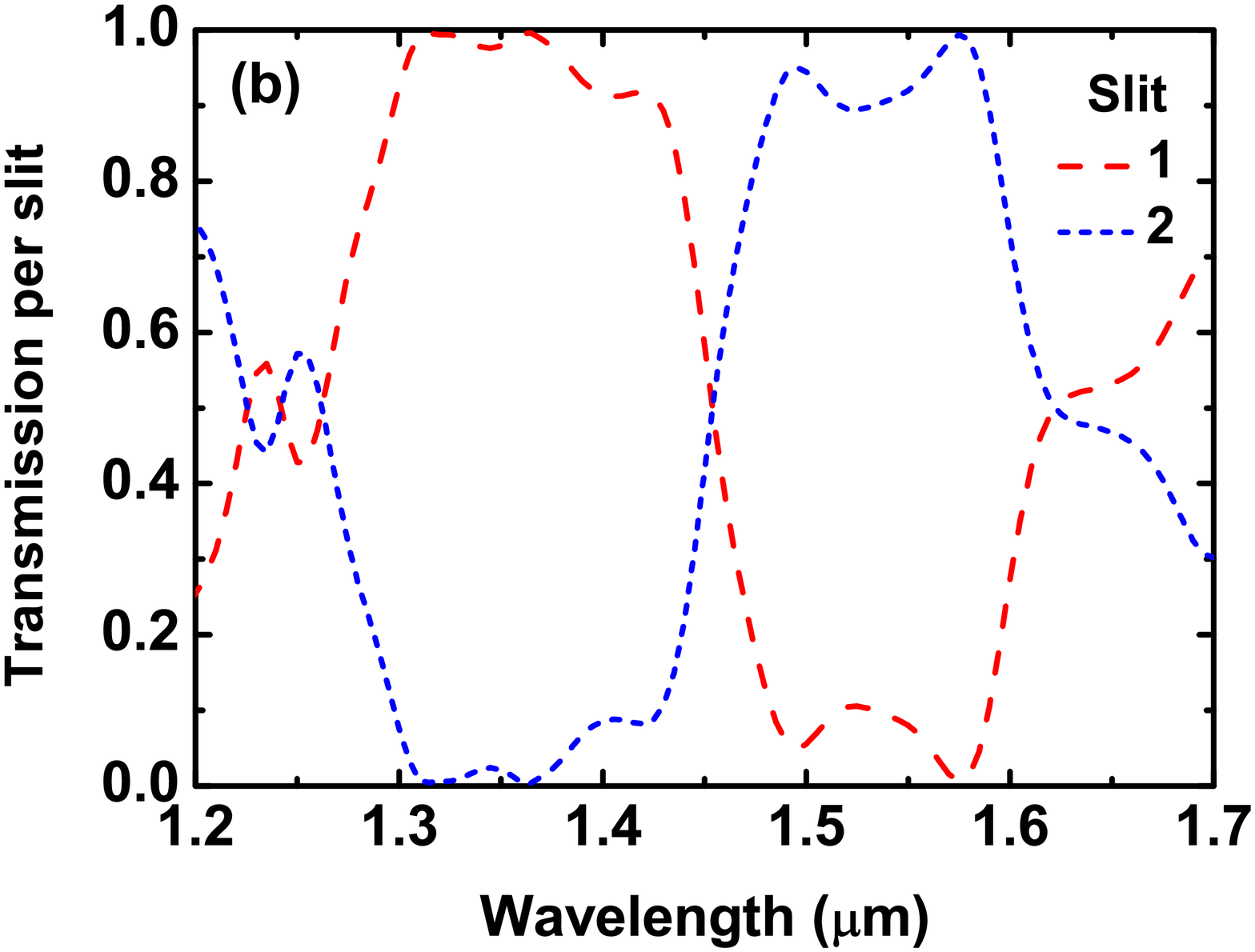}
\caption{(Color online). (a) Normalized-to-area transmittance ($\eta$) as a function of wavelength for the Double-Pixel (black-solid line) and the isolated pixels resonant to $\lambda_1=1.35 \mu$m (red-dashed line) and to $\lambda_2=1.5 \mu$m (blue-dotted line). The geometry of pixel 1 is $P_1=1236$ nm, $h_{g_1}=115$ nm, $w_{g_1}=363$ nm, and $d_{sg_1}=1135$ nm, while for pixel 2 we have $P_2=1380$ nm, $h_{g_2}=135$ nm, $w_{g_2}=363$ nm, and $d_{sg_2}=1280$ nm. The slit-slit distance of the Double-Pixel is $D_s=19.4$ $\mu$m. (b) Transmission per slit.}
\label{fig:spec}
\end{figure}

Let us first consider the optical response of a Double-Pixel with a given slit-slit distance $D_s=19.4$ $\mu$m, so that the constituent SGAs do not overlap. Fig \ref{fig:spec}(a) shows the normalized-to-area transmittance as a function of the wavelength. Optimal geometric parameters for both groove arrays are given in the caption of Fig. \ref{fig:spec}(a).

The Double-Pixel spectrum is compared with those for the isolated pixels 1 and 2, which exhibit narrow well-defined peaks with similar intensities. Peaks of the Double-Pixel are well resolved with a FWHM of about $100$ nm and a cross talk smaller than 1.0\%. The crosstalk is defined as the fraction of the total light transmitted by one pixel when only the other pixel is illuminated. 

Notice that the transmittance of the Double-Pixel is normalized to the power incident on the total area occupied by both slits. In order to compare the Double-Pixel on an equal footing with the isolated pixels, the spectra for the isolated pixels is divided by 2. That is equivalent to have a Double-Pixel with its constituent pixels  separated by an infinite distance. 

It is also worth to notice that the SGA in the isolated pixel 2 has a dip with vanishing $\eta$ at  $\lambda_1=1.35$ $\mu$m, see the blue-dotted line in Fig. \ref{fig:spec}(a). That explains the weak interaction between the two pixels at $\lambda_1$, reported in the next section. On the other hand, the intensity of the SGA in the isolated pixel 1 decays when the system is off resonance, but  still has a non-vanishing intensity at $\lambda_2=1.50$ $\mu$m, see the red-dashed line in Fig. \ref{fig:spec}(a), leading two an optical interaction between the two pixels at this wavelength. 

Fig \ref{fig:spec}(b) shows the transmission per slit, which is defined as the ratio of the EM power computed inside each slit and the total transmitted power. We can see that photons with wavelength $\lambda_1$ are mainly redirect to the slit 1, while most $\lambda_2$ photons pass through the slit 2. It means that the Double-Pixel behaves as an efficient photon sorter.

We find that the photon sorting in the Double-Pixel strongly depends on the relative position of the pixels, which is characterized by the distance between the slits ($D_s$). This behavior is discussed in the next section.

\subsection{Dependence on the slit-slit distance}
\label{sec:sorting}

\begin{figure}[h]
\centering
\includegraphics[width=8cm]{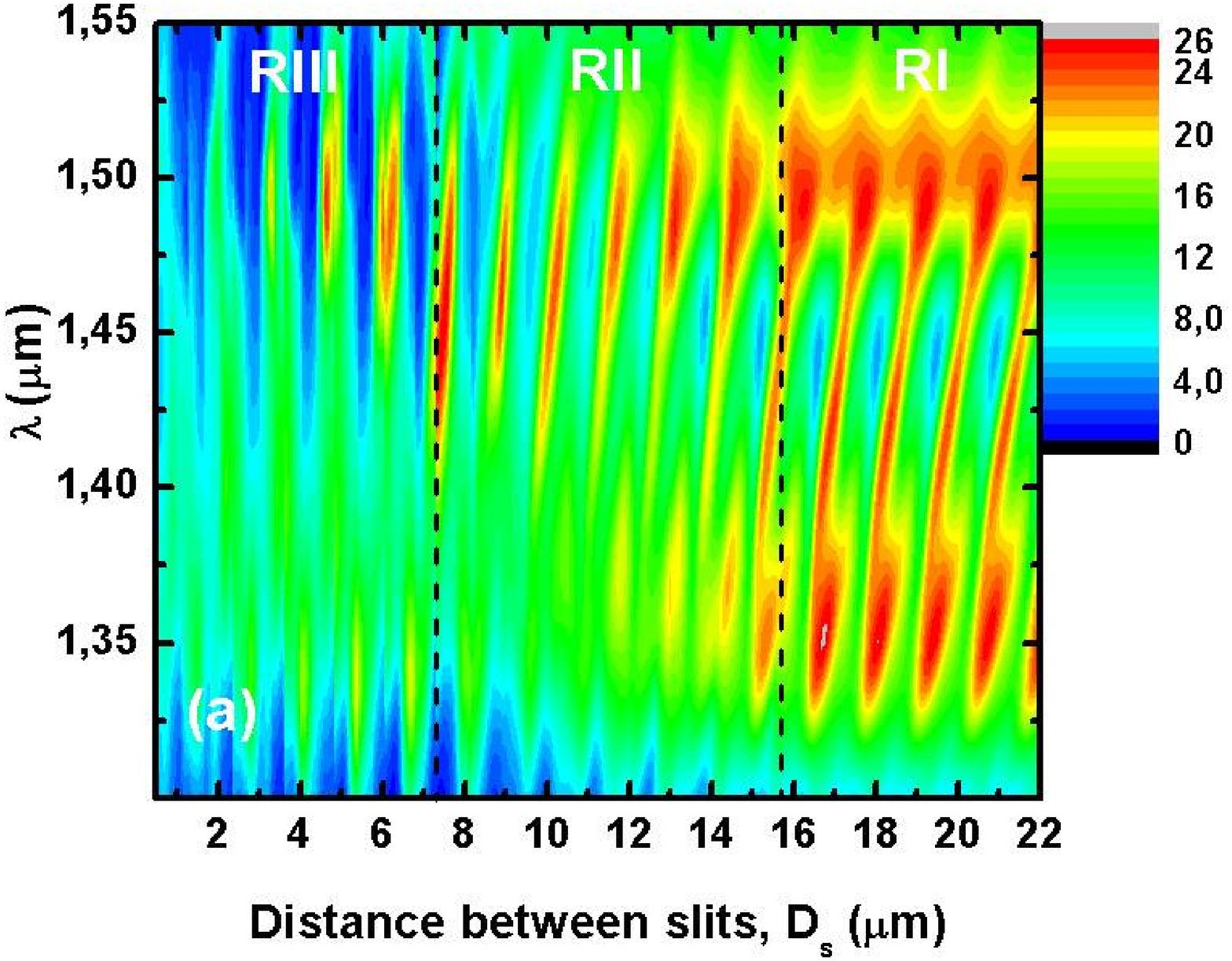}
\includegraphics[width=8cm]{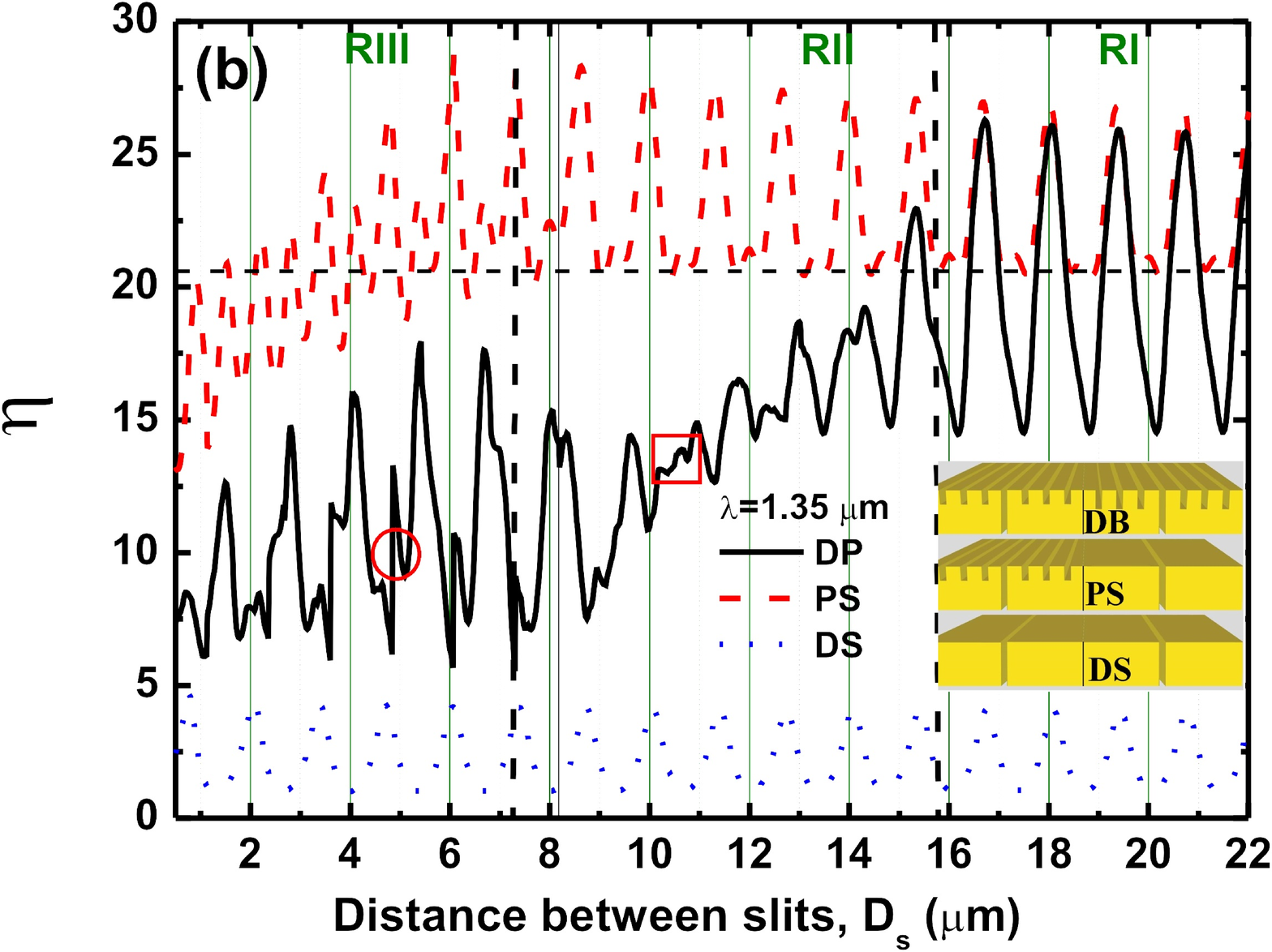}
\includegraphics[width=8cm]{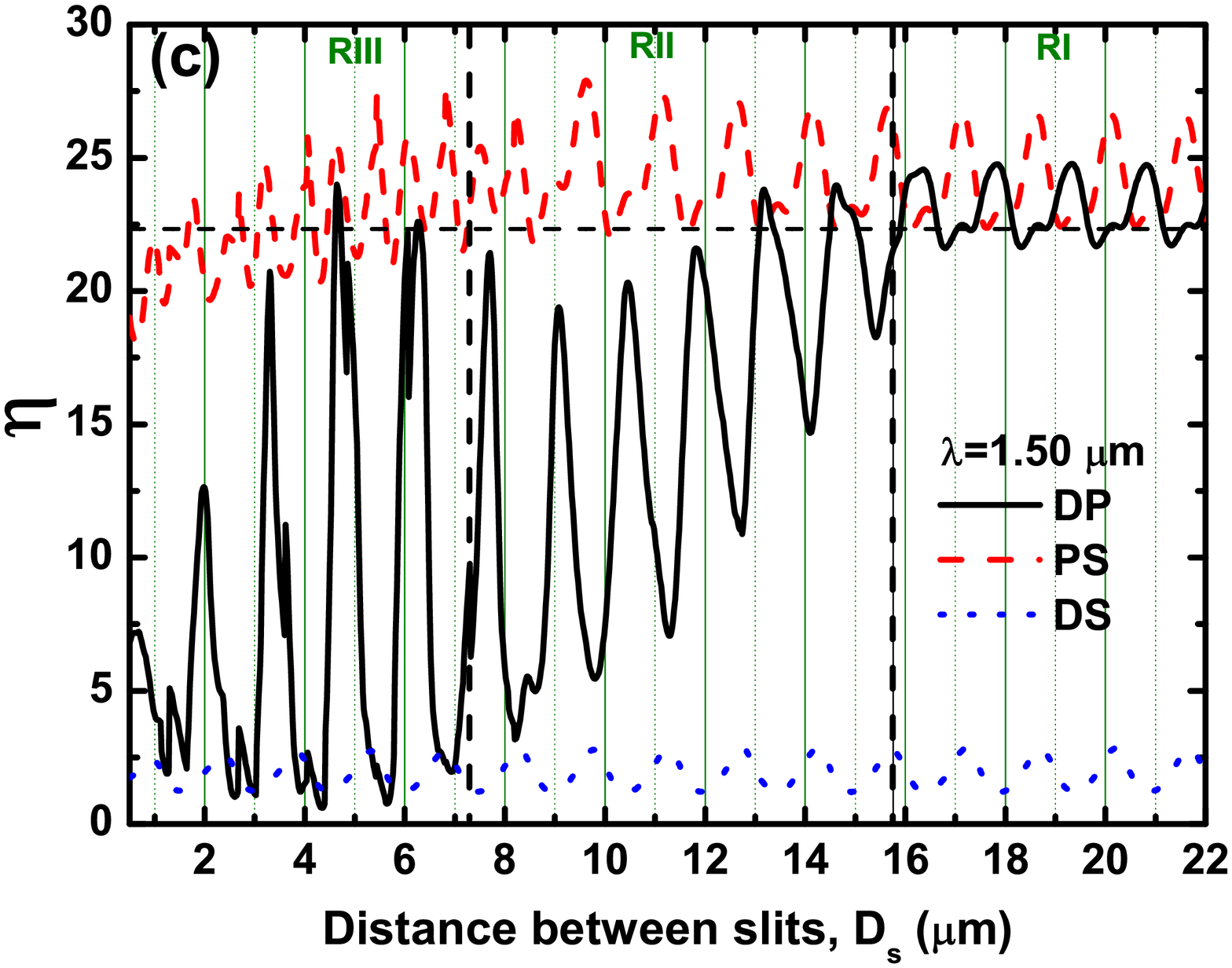}
\caption{(Color online). (a) Contour plot of $\eta$ for the Double-Pixel as a function of the wavelength $\lambda$ and the slit-slit distance $D_s$. (b) Crosscut at $\lambda_1=1.35$ $\mu$m for three different systems: i) Double-Pixel (DP, black line), ii) Pixel-Slit (PS, red-dashed line), which consists in pixel 1 and a single slit $S_2$, and iii) Double-Slit (DS, blue-pointed line). The inset shows schemes of the three structures. (b) Crosscut at $\lambda_1=1.50$ $\mu$m for DP, PS and DS systems. Geometrical parameters of slit and grooves are the same than in Fig. \ref{fig:spec}(a) (b).}
\label{fig:cp}
\end{figure}

Fig. \ref{fig:cp} illustrates the dependence of the transmission spectrum on the slit-slit distance. The contour plot in Fig. \ref{fig:cp} (a) represents $\eta$ as a function of both  wavelength and $D_s$. Crosscuts at $\lambda_1=1.35$ $\mu$m and $\lambda_1=1.50$ $\mu$m are shown in Fig. \ref{fig:cp} (b) and (c), respectively. 

Three different regimes are distinguished based on the effective area shared by overlapping pixels:
\begin{enumerate}
 \item[i)] In regime I (RI), the two pixels do not overlap. It occurs in Fig. \ref{fig:cp} for $Ds>15.7$ $\mu$m.
 \item[ii)] In regime II (RII), the grooves of one pixel overlap the neighbor groove array but without reaching the slit of the second pixel. It comprises the interval  $8.2$ $\mu$m $<D_s<15.7$ $\mu$m in Fig. \ref{fig:cp}. 
 \item[iii)] In regime III (RIII), grooves of one pixel overlap the slit of the neighbor pixel. See interval $D_s<8.2$ $\mu$m in Fig. \ref{fig:cp}.
\end{enumerate}

Before analyzing the physical trends observed in each of the regimes, it is worth to describe how we build the structure when two objects overlap. We discuss first the case of overlapping grooves and next  the case that the slit overlaps with the grooves.

When the grooves overlap, we have tried several rules for building the system. Larger transmission intensities are obtained when the grooves are separated an optimal edge-to-edge distance, which is centered at the initial midpoint, and the depth of the two overlapping grooves is the same than other grooves in the pixel. The ideal edge-to-edge distance is found to be of $20$ nm for the system considered here. Other two less efficient rules have been considered: (i) the two overlapping grooves are replaced by a single wider groove, (ii) one groove is kept fixed while the other is shifted to a nearby non-overlapping position. These additional rules are not discussed in the paper for they provide a poorer response of the system. 

When the slit overlaps with a groove of the neighbor pixel, the groove is displaced an edge-to-edge distance of $20$ nm from the slit. The reason for moving only the groove is explained below.

Let us to analyze now the three regimes observed in Fig. \ref{fig:cp} starting with regimen I. Transmission peaks in RI are repeated periodically with a distance between slits that is a multiple of the SPP wavelength ($\lambda_{spp1}=1.34$ $\mu$m and $\lambda_{spp2}=1.49$ $\mu$m for $\lambda_1$ and $\lambda_2$, respectively). Fig. \ref{fig:cp} (a) shows that peaks at fixed wavelengths $\lambda_1$ and $\lambda_2$ alternate their positions as a function of $D_s$. In fact, $\eta$ at the distance $D_s=19.4$ $\mu$m, used above in Fig. \ref{fig:spec}(a), is between the two local maxima for $\lambda_1$ and $\lambda_2$, in order to obtain similar intensities for the double-peak in the Double-Pixel. Such behavior is related to the different physical origin of the peaks, which is better illustrated in Figs. \ref{fig:cp} (b) and (c). 

We first fix the wavelength at $\lambda_1=1.35$ $\mu$m and vary $D_s$, as shown in  Fig. \ref{fig:cp} (b). In order to explain the optical response of the Double-Pixel, it is compared with two simpler systems: a Double-Slit (DS) for which all grooves are removed and only a single slit remains in each pixel (see \cite{HibbinsAPL02,SoenninchensenAPL00,SchoutenPRL05,LalannePRL05,AlarverdyanNP07,PacificiOE08,FdLPPRB11,HafeleAPL12} for a full discussion of its optical response), and the Pixel-Slit (PS) structure, where the grooves of one pixel are removed leaving only a single slit, while the other pixel is not changed. A schematic representation of the three systems is given in the inset of Fig. \ref{fig:cp} (b). 


The main peaks of the three systems are at the same spectral positions, c.f$.$ Double-Pixel, Pixel-Slit and Double-Slit curves in RI of Fig. \ref{fig:cp} (b). Therefore, the origin of such peaks can be attributed to the interaction between the two slits as in the simplest Double-Slit structure. However, the intensity of the peaks of both the Double-Pixel and the Pixel-Slit is 5 times larger than for the Double-Slit one. This is due to the presence of the groove array in pixel 1, which acts like an antenna to couple the incident light into surface modes that squeeze the EM energy into the central aperture of this pixel. 

Moreover, the interaction of the SGA in pixel 1 and the single slit in pixel 2 of the Pixel-Slit provides practically the same intensity than for the Double-Pixel, c.f.  red-dashed and black-solid lines in Fig. \ref{fig:cp} (b). Thus, the interaction with the groove array of pixel 2 can be, in principle, neglected. In the Pixel-Slit spectrum we observe small secondary peaks due to the interaction of the slits with the groove array in pixel 1. Such peaks are transformed into either small shoulders or asymmetric peaks in the Double-Pixel. 

In contrast, peaks of the Double-Pixel excited at $\lambda_2=1.50$ $\mu$m can be related with secondary peaks of the Pixel-Slit, see RI in Fig. \ref{fig:cp} (c). Therefore, the interaction between the two groove arrays can not be neglected in this case. 

Such different trends in Figs. \ref{fig:cp} (b) and (c) can be better understood looking back at the analysis of Fig. \ref{fig:spec}(a). We recall that the weak interaction between the two pixels at $\lambda_1$ is related to the minimum in the spectra of pixel 2 at this wavelength, while the optical interaction between the two pixels at  $\lambda_2$ is due to the tail in the peak of pixel 1. 

In order confirm our predictions, we have reduced in 40 nm the distance between grooves in pixel 1 (originally optimized at $\lambda_1=1.35$ $\mu$m), increasing in this way the spectral separation between the two peaks, and observed that peaks of the Double-Pixel excited at $\lambda_1=1.50$ $\mu$m moves to values of $D_s$ at which are excited the main peaks of the Pixel-Slit, as in Fig. \ref{fig:cp} (a), (such calculations are not shown in the paper). The behavior of the secondary peaks become more relevant in regimes II and III described in what follows.

As the two pixels approach each other and the groove arrays overlap, their stronger interaction produces an ``anticrossing`` of the two resonances, see Fig. \ref{fig:cp} (a). The nearest spectral separation between the two peaks is found at the boundary between RII and RIII. The effect of the anticrossing in RII is that the highest intensities  are no longer at the targeted wavelengths $\lambda_1=1.35$ $\mu$m and $\lambda_2=1.50$ $\mu$m. The peak at $\lambda_1$ is red shifted when the distance between the slits is reduced, while the peak at $\lambda_2$ is blue shifted. Peaks became also narrower than in RI and their relative intensities change so that peaks at $\lambda_1$ have lower intensities than those at $\lambda_2$. 

Fixing $\lambda_1=1.35$ $\mu$m, as in in RII of Fig. \ref{fig:cp} (b), we observe not only  the aforementioned reduction of the transmitted intensity, but also a departure of the Double-Pixel response from the behavior associated to a Pixel-Slit. Furthermore, secondary peaks of the Pixel-Slit (related with the slit-groove interaction) becomes more relevant for the Double-Pixel, while peaks related to the slit-slit interaction in the Pixel-Slit are strongly suppressed by the new  conditions of interference. Such effects become more pronounced as the slits approach each other. Similar features are found for  $\lambda_2=1.50$ $\mu$m, see Fig. \ref{fig:cp} (c). The main difference with Fig. \ref{fig:cp} (b) is that secondary peaks have been already excited in RI and only become better defined in RII, though their intensity also decrease for the presence of the anticrossing. 

Additional minima appear when grooves of different pixels occupy the same region of the space, see for instance the interval demarcated by a red square in Fig. \ref{fig:cp} (b), where more than 60\% of the grooves of one pixel overlap with grooves of the other pixel. 

When the slits enter into the overlapping region, as in RIII of Fig. \ref{fig:cp}, resonances move away the anticrossing point and the intensity of the transmission peaks starts to raise.  Peaks in RIII become narrower and better defined than in RII.  In particular, peaks at $\lambda_2=1.50$ $\mu$m practically reach  the intensity of RI. This feature is useful for applications for the system in RIII covers an area smaller than in RI. 

When the slit occupies the position that would correspond to a groove of the neighbor pixel, we find that the slit transmits additional light leading to the secondary narrow peaks demarcated by a red circle in Fig. \ref{fig:cp} (b).  Such narrow peaks have the same spectral position than peaks related to the slit-slit interaction in the Pixel-Slit. The building rule, defined at the beginning of the section for the case of overlapping slit and grooves, takes advantage of this feature.

\subsection{Influence of the number of grooves} 
\label{sec:number}
\begin{figure}
\includegraphics[width=8cm]{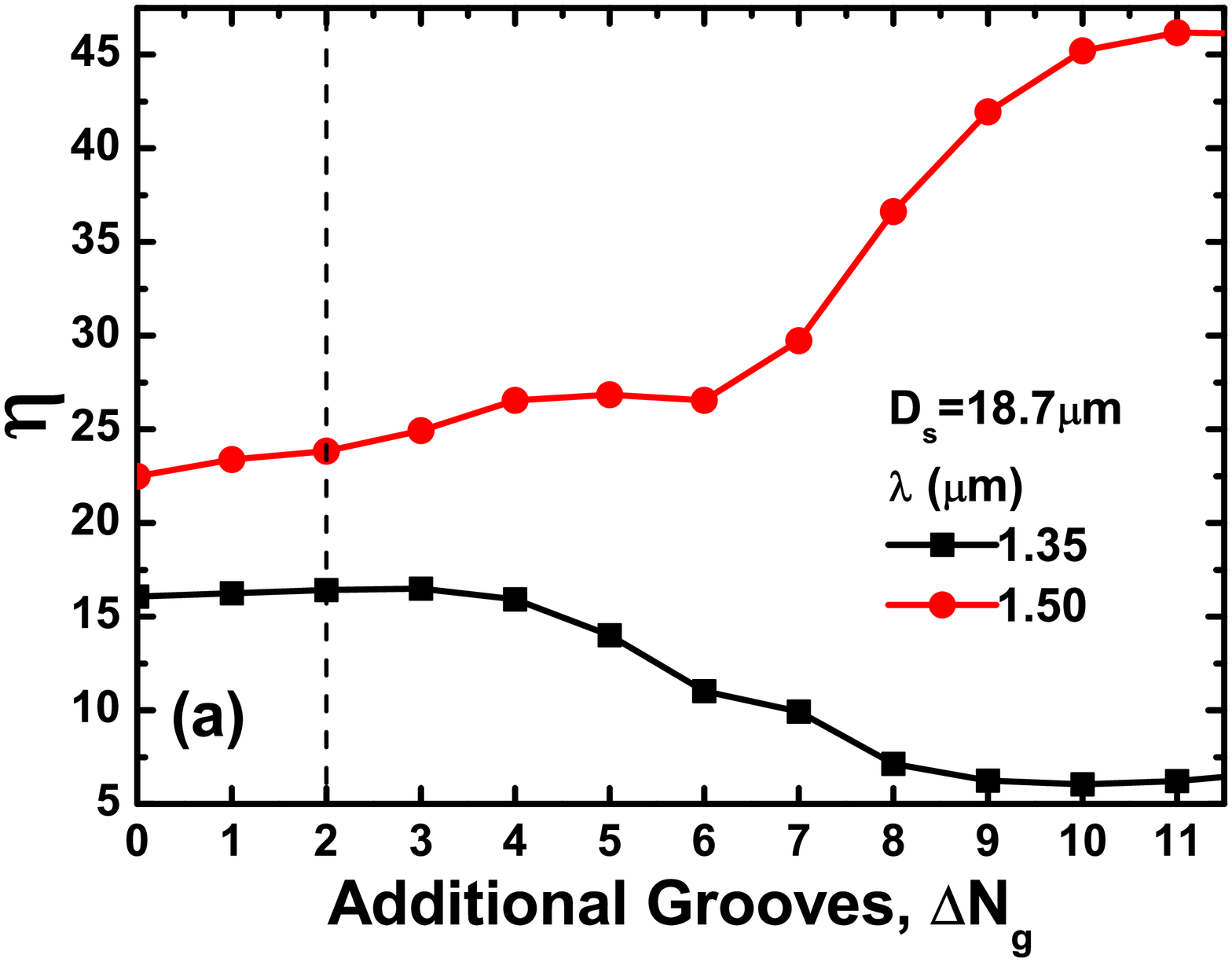}
\includegraphics[width=8cm]{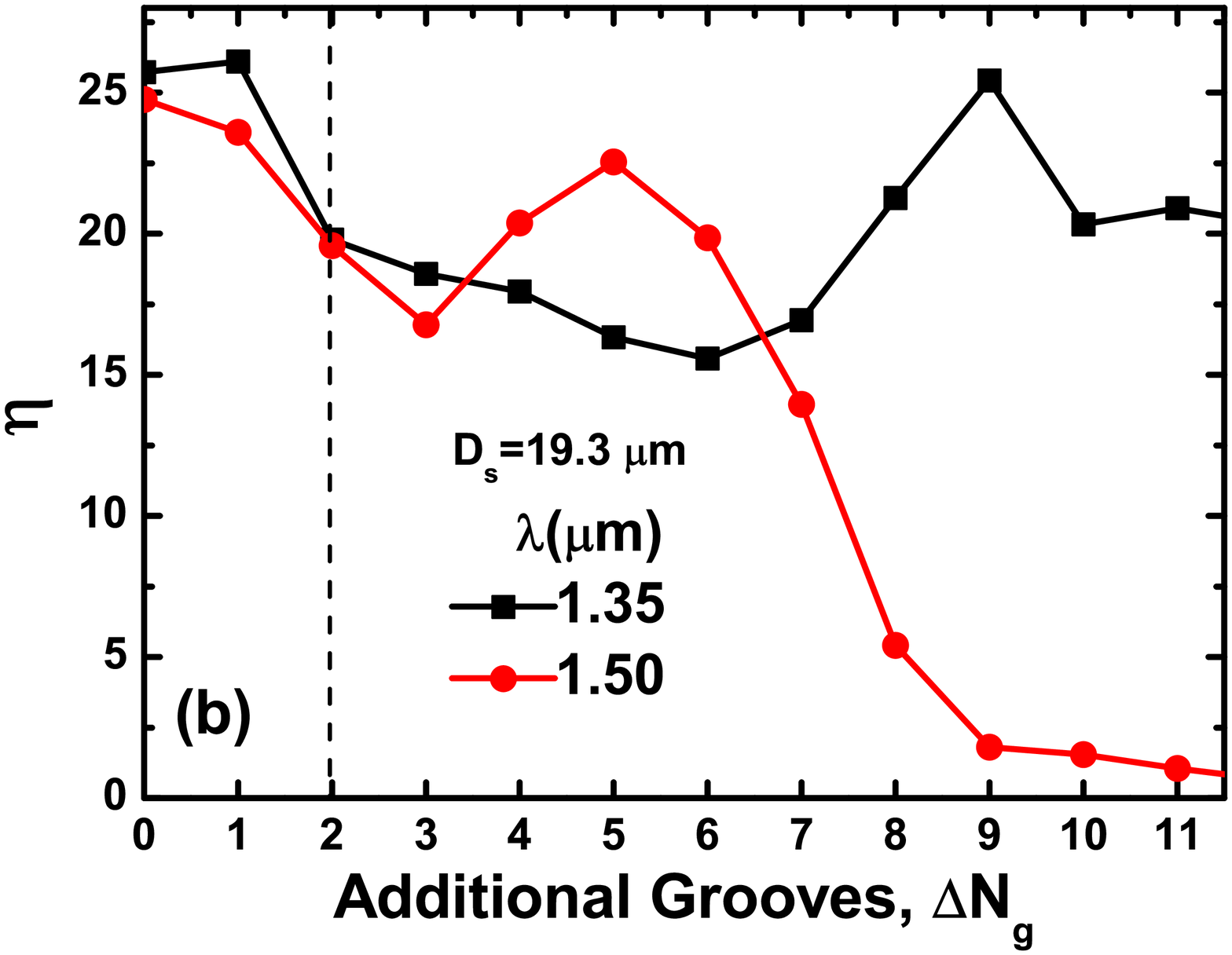}
\caption{(Color online). Intensity of the transmittance peaks for the two targeted wavelengths ($\lambda_1=1.35$ $\mu$m and $\lambda_2=1.50$ $\mu$m) as function of the number of grooves added to: (a) pixel 2 ($\Delta N_g \equiv N^L_{g_2}-N^R_{g_2} \geqq 0$ and $\Delta N_{g_1}=N^R_{g_1}-N^L_{g_1}=0$, see definition of the structure in Fig. \ref{fig:scheme}) and (b) both pixels ($\Delta N_g \equiv N^R_{g_1}-N^L_{g_1}=N^L_{g_2}-N^R_{g_2} \geqq 0$). Geometrical parameters of slits and grooves are the same than in Fig. \ref{fig:spec}. The vertical line represents the minimal $\Delta N_g$ to have overlapping pixels.}
\label{fig:sort}
\end{figure}

It is also worth to study the photon sorting as a function of the number of grooves. We start with the Double-Pixel already optimized in Fig. \ref{fig:cp}, which has 12 grooves in each pixel. Our goal is to increase the number of grooves that redirect additional photons through the apertures, but keeping fixed the total size of the system. Thus, the grooves are added between the two pixels, either to the right of pixel 1 or to the left of pixel 2, see Fig. \ref{fig:scheme}.
\\
As a proof of principle of the photon sorting, we consider first the situation in which only grooves with the same geometrical parameters as in pixel 2 are added to the left of this pixel. The two SGAs are separated a distance $D_s=18.7$ $\mu$m, for  which the system is resonant at $\lambda=1.5$ $\mu$m, see Fig. \ref{fig:cp}(c). The intensity of the transmittance peaks as a function of the additional grooves is represented in Fig. \ref{fig:sort}(a).   We find an enhancement of the intensity of the the peak at $\lambda=1.5$ $\mu$m and a concomitant reduction of the intensity at $\lambda=1.35$ $\mu$m. 

Taking also into account that the two targeted wavelengths are excited at different values of $D_s$, see Fig. \ref{fig:cp}, we conclude that is not possible to simultaneously enhance the efficiency of photon sorting for both kind of photons. According with our calculations, this physical constrain can not be overcome even optimizing each additional groove independently or implementing chirped groove arrays as in Ref. \cite{LauxNP08}. 

As a rule of thumb, we suggest to use a moderate number of additional grooves. A typical case is illustrated in Fig. \ref{fig:sort} (b), where the two pixels are separated the same distance $D_s=19.4$ $\mu$m than in Fig. \ref{fig:spec}(a)(b). We observe a systematic reduction of $\eta$ with $\Delta N_g$ and local maxima for different number of additional grooves ($\Delta N_g=5$ and $\Delta N_g=9$ for $\lambda_1=1.35$ $\mu$m and $\lambda_2=1.50$ $\mu$m, respectively). So, the intensity is large enough for both wavelengths when $\Delta N_g \leq 7$. Notice that, despite this reduction in intensity, the overlap of the two pixels is still convenient for practical applications due to the reduction of the total size of the system, as already pointed out in Ref. \cite{LauxNP08}. 

\section{Conclusions}
We have studied the process of sensing and sorting photons with different wavelengths  by a Double-Pixel. Each pixel consists of a slit-groove array optimized to harvest light of a given wavelength.

 We find that the optical interaction between the slit-groove arrays strongly depends on distance between slits. Three different regimes for the process of photon sorting are identified: (i) non-overlapping pixels, (ii) pixels where only grooves are overlapped, and (iii) pixels where grooves also overlap with the slits. 

The spectral position of the two resonant peaks approaches an anticrossing point when the groove arrays of the two pixels overlap each other. A reduction of the size of the system due to the overlapping of the pixels does not impair the transmission efficiency. In fact, the intensity of the transmittance peaks can be so large as for non-overlapping pixels  when slits enter into the overlapping region. A moderate number of grooves is needed for efficiently photon sorting at two different wavelengths.


 Similar mechanisms are expected for the 2D version of the photon sorter (the bull's eye geometry studied in Ref. \cite{LauxNP08}), though a detailed study of this more involved structure exceeds the goals of the present paper. Therefore, we hope that the present  study could motivate further experiment and theoretical works, and pave the way for future applications.

\ack{The authors gratefully acknowledge financial support by European Projects EC FP7-ICT PLAISIR Project  Ref. 247991 and the Spanish Ministry of Science and Innovation project MAT2011-28581-C02-02.}

\section*{References}
\bibliographystyle{njpstyle4}
\bibliography{ref}


\end{document}